\documentclass[12pt]{article}
\usepackage{graphicx}
\usepackage{dcolumn}
\usepackage{bm}
\usepackage{color}
\usepackage{amssymb}
\usepackage{amsmath}
\usepackage{natbib}
\usepackage{multicol,times}
\usepackage{multirow}
\usepackage{mathrsfs}
\usepackage{url}

\begin{document}

\title{\bf A Consistent Model of `Explosive' Financial Bubbles With Mean-Reversing Residuals}
\author{ L. Lin{\small$^{\mbox{\,\ref{ETH}, \ref{BeiHang}~\footnote{Email: llin@ethz.ch}}}$} ,
\, \, Ren R.E.{\small$^{\mbox{\,\ref{BeiHang}}}$} and \, D. Sornette{\small$^{\mbox{\,\ref{ETH}, \ref{SFI}
\footnote{Email: dsornette@ethz.ch}}}$}
}%
\date{}

\maketitle

\vspace{-9mm}

\begin{enumerate}
\item{Chair of Entrepreneurial Risks, Department of Management, Technology and Economics, ETH Zurich, Kreuplatz 5, CH-8032 Zurich, Switzerland}\label{ETH} \vspace{-3mm}
\item{School of Economics and Management, Beihang University, 100191 Beijing, China}\vspace{-3mm}
\label{BeiHang}%
\item{Swiss Finance Institute, c/o University of Geneva, 40 blvd. Du Pont d'Arve, CH 1211 Geneva 4, Switzerland}\label{SFI}%
\end{enumerate}

\vspace{1mm}
\begin{center}
({\em{version May 1, 2009}}) 
\end{center}
\vspace{1mm}
\begin{center}
\textbf{Abstract}
\end{center}

\begin{quote}
\hspace{0.5cm} 
We present a self-consistent model for explosive financial bubbles, 
which combines a mean-reverting volatility process
and a stochastic conditional return which reflects nonlinear positive feedbacks
and continuous updates of the investors' beliefs and sentiments.
The conditional expected returns
exhibit faster-than-exponential acceleration
decorated by accelerating oscillations, called ``log-periodic power law.''
Tests on residuals show a remarkable low rate ($0.2\%$) of
false positives when applied to a GARCH benchmark. When tested on the S\&P500 US index
from Jan.~3, 1950 to Nov.~21, 2008, the model correctly identifies the bubbles
ending in Oct.~1987,  in Oct.~1997, in Aug.~1998 and 
the ITC bubble ending on the first quarter of 2000. Different unit-root tests confirm
the high relevance of the model specification. Our model also provides a diagnostic
for the duration of bubbles: applied to the period before Oct. 1987 crash, there is clear
evidence that the bubble started at least 4 years earlier. 
We confirm the validity and universality of the volatility-confined LPPL model 
on seven other major  bubbles that have occurred in the World in the last two decades.
Using Bayesian inference, 
we find a very strong statistical preference for our model compared with a standard
benchmark, in contradiction with \citet{Feigenbaum2006} which used
a unit-root model for residuals.

\end{quote}

\vspace{0.5mm}

\begin{quote}
\end{quote}

\begin{quote}
\end{quote}

\section{Introduction}\label{s:intro}

We present a self-consistent model for explosive financial bubbles, 
with nonlinear positive feedbacks
with mean-reversal residuals. The conditional expected returns
exhibit faster-than-exponential acceleration
decorated by accelerating oscillations.
An essential advance of our model compared with previous specifications
such as that of Johansen-Ledoit-Sornette (1999)\nocite{JLS1} 
is to allow for stochastic conditional expectations of returns which 
describe continuous updates of the investors' beliefs and sentiments. 

Two different modeling
strategies lead to the same final model specification: (i) a rational-expectation (RE) model
of rational bubbles with combined Wiener and Ornstein-Uhlenbeck innovations describing
the dynamics of rational traders coexisting with noise traders driving the crash hazard rate;
or (ii) a behavioral specification of the dynamics of the 
stochastic discount factor describing the overall combined decisions of both rational
and noise traders. 

Tests on residuals show a remarkable low rate ($0.2\%$) of
false positives when applied to a GARCH benchmark. When tested on the S\&P500 index
from Jan. 3, 1950 to Nov. 21, 2008, the model correctly identifies the bubbles
ending in Oct. 1987,  in Oct. 1997 and in the summer of 1998 and 
the ITC bubble ending on the first quarter of 2000. Different unit-root tests confirm
the high relevance of the model specification. Our model also provides a diagnostic
for the duration of bubbles: applied to the period before Oct. 1987 crash, there is clear
evidence that the bubble started at least 4 years earlier. Using Bayesian inference, 
we find a very strong statistical preference for our model compared with a standard
benchmark, in contradiction with the result of \citet{Feigenbaum2006}. Our 
positive result stems from the mean-reverting structure of the residuals
of the conditional returns modeling the bubbles, which is shown to be
essential in order to obtain a consistent model. 
Absent in previous specifications, this feature constitutes the main advance
of this work, leading to the novel positive results.
The same tests performed on seven major bubbles 
(Hong Kong 1997,  ITC 2000 bubble, Oil bubble ending July 2008, the Chinese bubble
ending in October 2007 and others) suggest that our
proposed volatility-confined LPPL model provides a consistent universal description of financial bubbles, namely
a super-exponential acceleration of price decorated with log-periodic oscillations
with mean-reverting residuals.

The present work offers an innovative way to break the stalemate in the 
ex-ante detection of bubbles, which has been much discussed
in the literature. For instance,  \citet{Gurkaynak2008} summarizes econometric approaches applied
to the detection of financial bubbles, stating that the ``econometric detection of asset price bubbles cannot be achieved with a satisfactory degree of certainty. For each paper that finds evidence of bubbles, there is another one that fits the data equally well without allowing for a bubble. We are still unable to distinguish bubbles from time-varying or regime-switching fundamentals, while many small sample econometrics problems of bubble tests remain unresolved.'' 

Bubbles are often defined as exponentially explosive prices, which are followed by a sudden collapse. As summarized for instance by  \citet{Gurkaynak2008}, the problem with this definition is that any exponentially growing price regime, that one would call a bubble, can be also rationalized by a fundamental valuation model. This is related to the problem that the fundamental price is not directly observable, giving no indisputable anchor to understand how observed prices may deviate from fundamental values. This was exemplified during the last Internet bubble culminating in 2000 by fundamental pricing models, which incorporated real options in the fundamental valuation, basically justifying any price. \citet{Michael1999} were among the most vocal proponents of the proposition offered close to the peak of the Internet bubble, that better business models, the network effect, first-to-scale advantages, and real options effect could account rationally for the high prices of dot-com and other New Economy companies. These interesting views expounded in early 1999 were in synchrony with the general positive sentiments of the bull market of 1999 and preceding years. They participated in the general optimistic view and added to the strength of the herd. Later, after the collapse of the bubble, these explanations seemed less attractive. 

Our model addresses in an innovative way this problem of defining and identifying bubbles.
It extends in a novel direction a class of processes that have been proposed to incorporate the 
positive feedback mechanisms that can push prices upward faster-than-exponentially. This
faster-than-exponential characteristics is one of the main diagnostic that we consider for a bubble.
Many financial economists recognize that positive feedbacks and in particular herding is a key factor for the growth of bubbles. Herding can result from a variety of mechanisms, such as anticipation by rational investors of noise traders strategies \citep{Delong1990}, agency costs and monetary incentives given to competing fund managers \citep{Dass2008} sometimes leading to the extreme Ponzi schemes \citep{Dimi2004}, rational imitation in the presence of uncertainty \citep{Didier2000}, and social imitation. The relevance of social imitation or ``word-of-mouth'' effects has a long history (see for instance [\citet{Shiller2000, Hong2005}]
for recent evidence). Our approach is to build on previous specifications that describe
faster-than-exponential growth of price (coined hereafter ``super-exponential'') \citep*{Didier2002, Didier2003}. 

The Johansen-Ledoit-Sornette (JLS) model \citep{JLS1,JLS2} constitutes a first attempt to formulate these ingredients into a traditional asset pricing model. Starting from the rational expectation model of bubbles and crashes developed by \citet{blanchard1979} and by \citet{blanchard1982}, the JLS model considers the critical properties inherent in the self-organization of complex systems. In the JLS model, the financial market is composed of two types of  investors: perfectly rational investors who have rational expectations and irrational traders who are prone to exhibit herding behavior. The dynamics of the price is described by the usual geometric Brownian motion plus a jump process controlled by its crash hazard rate. The noise traders drive the crash hazard rate according to their collective herding behavior, leading its critical behavior. Due to the no-arbitrage condition, this is translated into a price dynamics exhibiting super-exponential acceleration, with possible additional so-called ``log-periodic'' oscillations associated with a hierarchical organization and dynamics of noise traders. Using the stochastic discount factor (SDF), \citet{Weixing2006} extended the JLS model to include inter-temporal parameters and  fundamental economic factors.

 In the Johansen-Ledoit-Sornette (1999, 2000) model,  the
logarithmic return is drawn from a normal distribution with a time-varying drift,
\begin{equation}\label{E: r_JLS}
r_{i}=\ln p_{t_{i+1}}-\ln p_{t_i}\,\sim\,N(\Delta H_{t_{i+1},t_{i}},\,\sigma^2 (t_{i+1}-t_i)), \quad \Delta H_{t_{i+1},\, t_{i}}=H_{t_{i+1}}-H_{t_{i}}~,
\end{equation}
where
\begin{equation}\label{E: landau 1}
H_{t_{i}}=A-B(t_c-t_{i})^{\beta}\left[1+\frac{C}{\sqrt{1+(\frac{\omega}{\beta})^2}}\cos(\omega\ln (t_c-t_{i}))+\phi)\right]~.
\end{equation}
This so-called log-periodic power law (LPPL) dynamics given by (\ref{E: landau 1}) has been previously
proposed in different forms in various papers  (see for instance
\citet{Didier1996, Feigenbaum1996,  Didier1999, Didier2001, Feigenbaum2001, Weixing2003, Drozdz2003, Whycrash}).
The power law $A-B(t_c-t_{i})^{\beta}$
expresses the super-exponential acceleration
of prices due to positive feedback mechanisms,
alluded to above. Indeed, for $B>0$ and $0 < \beta < 1$, the rate of change of $H_{t_{i}}$
diverges as $t \to t_i^-$.
The term proportional to $\cos(\omega\ln (t_c-t_{i}))+\phi)$ describes
a correction to this super-exponential behavior, which 
has the symmetry of {\em discrete scale invariance (DSI)} \citep{Didier1998}.
This formulation (\ref{E: landau 1}) results from analogies with critical phase transitions (or bifurcations) occurring
in complex adaptive systems with many interacting agents. The key insight is that
spontaneous patterns of organization between investors emerge from repetitive interactions at the micro-level,
possibly catalyzed by top-down feedbacks provided for instance by the media and macro-economic readings,
which are translated into observable bubble regimes and crashes. A common mathematical
signature of such critical behavior is found in the power law singularities that accompany
the faster-than-exponential growth. The additional acceleration oscillations may result
from the existence of a discrete hierarchy of the organization of traders \citep{Didier1998b}, or from the
interplay between the inertia of transforming information into decision together with
nonlinear momentum and price-reversal trading styles \citep{Ide}.

Previous tests of the LPPL model (\ref{E: r_JLS}) with (\ref{E: landau 1}) and 
its variants belong to the following three main types:
\begin{enumerate}
\item non-parametric tests of the super-exponential behavior and especially of the log-periodic
oscillatory structure applied to residuals of prices time series \citep{Weixing2002, Weixing2003,Weixing2003b};
\item nonlinear least-square fits of price and log-price time series \citep{Didier2001, Didier2001b, Weixing2008, Didier2009};
\item Bayesian methods applied to the time series of returns \citep{Feigenbaum2006}.
\end{enumerate}
Each type has limitations.
\begin{itemize}
\item Non-parametric approaches to the LPPL models have focused essentially on 
testing the statistical significance of the log-periodic component of price residuals
in bubble regimes ending with crashes. In themselves, they do not provide
complete tests of the LPPL model (\ref{E: r_JLS}) with (\ref{E: landau 1}) and 
its variants.

\item Calibrating directly price or log-price time series may produce spurious
high measures of goodness of fits (\citealt{Granger1974, Phillips1986}).
As a consequence of their non-stationarity, the goodness of fit may not reflect the
 properties of the underlying data generating process. Indeed, prices
 or log-prices are to a good approximation generated by non-stationary unit-root processes,
 obtained from the integration of stationary returns. Such integration mechanically
 reddens the spectrum, damping the high-frequency component of the time series,
 which may lead to the illusion that the generating process is deterministic.
 
 \item This problem has led \citet{Feigenbaum2001} and \citet{Feigenbaum2006}
to propose tests of the LPPL model applied to the return time series.  Indeed, the LPPL model 
(\ref{E: r_JLS}) with (\ref{E: landau 1}) also predict a LPPL structure for the returns.
The difficulty with this approach is that direct filters of the LPPL patterns 
from daily returns have been unable until now to detect a signal predicted to be one-order-of magnitude
smaller than the background noise (\citet{Feigenbaum2001}; see however \citet{Didier2001b}
for a more positive reinterpretation of Feigenbaum's results). The standard financial econometric response
to this problem is to work with monthly or quarterly time scales, so that the volatility is
reduced in relative value compared to the drift, approximately by the square root
of the number of days in a month or in a quarter.
Unfortunately, this is hardly applicable to the  problem of detecting and
calibrating financial bubbles since the 
signal we are looking for is by construction transient. Therefore, the luxury of 
long time series spanning many months or quarters is not available. If a bubble
expands over 4 years, this provides only 48 months and 16 quarters, not sufficient
to calibrate econometric models.
\citet{Feigenbaum2006} later made the first attempt to employ 
a Bayesian method which is better suited for the
analysis of complicated time-series models like the JLS
model expressed in terms of returns.
Through the comparison of marginal likelihoods, they discovered that, if they
did not consider crash probabilities, a null hypothesis model without log-periodical structure outperforms the JLS model. And if the JLS model was true, they found that parameter estimates obtained by curve fitting
have small posterior probability.  Even though the LPPL hypothesis might be correct,
they concluded that researchers should abandon the class of models in which the LPPL structure
is revealed through the expected return trajectory.
\end{itemize}

These problems can be fundamentally traced back to the fact that the JLS model
describes a deterministic time-varying drift decorated by
a non-stationary stochastic random walk component. 
In accordance with rational expectation, this predetermined deterministic price path is 
the unbiased expectation of a representative rational agent in the market, while the 
stochastic component describes the estimation errors. The problem is that 
the  stochastic random walk component is a variance-increasing process, so that 
the deterministic trajectory strays farther and farther away from the observable price path. 
This is the reason why direct calibration of prices are inconsistent with the 
estimation of the unbiased expectation of prices. And, as we shall demonstrate below,
this is also the reason for the lack of power of the Bayesian approaches applied
to the return time series.

In this context, the innovation of our approach is to modify the JLS model by a new specification of the residuals,
that makes the process consistent with direct price calibration, thus addressing the issues raised by
\citet{Granger1974} and \citet{Phillips1986}. In a nutshell, 
the realized observable price path during bubbles is attributed to  a deterministic LPPL component,
while the estimation errors by rational investors is modeled by a mean-reversal {\em Ornstein-Uhlenbeck (O-U)}\footnote{In discrete times, it  becomes an AR(1) process} process. While keeping
the structure of the model based on time-varying expectations of future returns, 
the daily logarithmic returns are no longer described by a deterministic drift decorated
by a Gaussian-distributed white noise. Instead, specifying a mean-reversal noise component, 
the no-arbitrage condition predicts that the expected returns become stochastic, which 
represents the on-going reassessment by investors of the future returns. 

Section 2 presents the new model, which we call the ``volatility-confined LPPL model'',
from two different perspectives, a first derivation 
based on rational expectation and an equivalent demonstration using
the stochastic discount function.  Section 3 presents
a first battery of empirical statistical tests. Applying direct calibrations of the new LPPL
specification to prices generated by GARCH$(p,q)$  processes show that the rate of false
positives in terms of the detection of bubble regimes is smaller than 0.2\%.
Using tests on residuals of the price calibration method applied to shrinking windows converging to the 
crash of October 1987, we are able to identify a clear bubble regime starting about
4 years before the crash occurred.  Section 4 implements the Bayesian analysis, extending
the approach of \citet{Feigenbaum2006} to our LPPL specification with O-U residuals.
The results show a very strong significance of the LPPL model versus a standard benchmark,
as the marginal likelihood calculated from the data within bubbles prior to the Oct. 1987 crash 
is about 150 times larger than that of models in which daily returns have no LPPL structure. 
Section 5 presents the results of the tests of section 3 to seven other major bubbles 
(Hong Kong 1997,  ITC 2000 bubble, Oil bubble ending July 2008, the Chinese bubble
ending in October 2007 and others) to confirm that our
proposed volatility-confined LPPL model provides a consistent universal description of financial bubbles, namely
a super-exponential acceleration of price decorated with log-periodic oscillations
with mean-reverting residuals.  Section 6 concludes.

\section{Volatility-confined LPPL model}\label{s:2}

Our volatility-confined LPPL model can be obtained in two ways: (i)
using the traditional economic framework of rational expectation and (ii)
on the basis of the  {\em Behavioral Stochastic Discount Factor (BSDF)}.
Although both derivations lead to the same specification, they provide 
different and complementary economic interpretations. In the following
two subsections, we present in turn these two derivations.

\subsection{Derivation based on the {\em Rational Expectation (RE)} condition \label{rederiv}}

Let us consider a financial market in which a regime shift occurs, changing from a 
standard GARCH process into a bubble phase. The price dynamics in the bubble
regime is assumed to be given by the following process. 
\begin{subequations}\label{E: RE model}
\begin{align}
\frac{dI}{I} &= {\mu}(t)dt+\sigma_Y dY +\sigma_W dW-\kappa dj ~, \\
dY &= -\alpha Y dt + dW~.
\label{E: Y of OU}
\end{align}
\end{subequations}
The symbol $I$ denotes the stock index or price of the asset and $W$ denotes the
standard Wiener process. The time-varying drift leading to the price acceleration
which is characteristic of a bubble regime is represented by $\mu(t)$
and the jump process $j$ takes the value zero before the crash and one afterwards. The constant $\kappa$ denotes the percentage price drops during a crash. The stochastic process $Y$ plays an important role in the model. For $0<\alpha<1$, $Y$ is an Ornstein-Uhlenbeck process, so that $dY$ and $Y$ are both stationary. As we shall see, this property ensures that the calibration of the LPPL model to the price time series is consistent, which was
not the case for the standard JLS model in the absence of $Y$.
Equation \eqref{E: Y of OU} describes a self-stabilization mechanism occurring in the market that confines 
the volatility to remain bounded during the bubble gestation all the way until 
the downward jump (or crash) occurs. For $\alpha=0$ or in absence of $Y$, the model recovers the original form of the price dynamics in the JLS model. The JLS model is therefore nothing but a special case of our model (\ref{E: RE model}) with (\ref{E: Y of OU}).
The corresponding version in discrete time of (\ref{E: RE model}) with (\ref{E: Y of OU}) reads
\begin{subequations}\label{E: RE model 2}
\begin{align}
\ln I_{t+1}-\ln I_{t} &={\mu}_t+\sigma_Y (Y_{t+1}-Y_{t})+\sigma_W \varepsilon_t-\kappa \Delta j_{t}~, \\
Y_{t+1}&=(1-\alpha)Y_{t}+\varepsilon_{t} ~,
\end{align}
\end{subequations}
where $\varepsilon_{t}\,\sim\,N(0,\,1)$. 

Let us assume that the dynamics of the Stochastic Discount Factor (SDF) satisfies:
\begin{equation}\label{E: SDF}
\frac{d\Lambda_t}{\Lambda_t}=-rdt-\rho_Y dY-\rho_W dW~.
\end{equation}
The factor $r$ quantifies the difference between the risk-free interest rate $r_f$ and the dividend 
growth rate $\delta$ ($r=r_f-\delta$). The terms $\rho_Y dY$ and $\rho_W dW$ amount to transforming the objective drift of the return
process into its corresponding risk-neutral version, via the no-arbitrage condition (\ref{htjotjgwg}) written below.
The SDF can be interpreted as the excess return over the spot interest that an asset must earn per unit of risk variance associated respectively with the two processes $Y$ and $W$.  Only these two
stochastic processes need to be considered in the dynamics of $\Lambda_t$ since any others
which are uncorrelated with $Y$ and $W$ do not contribute to the pricing of the assets considered here.
The SDF $\Lambda$ is the pricing kernel of the financial market, that reflects the risk-neutral probability measure in which the current intrinsic price of any asset is equal to the value of its expected discount future payoffs. 
When the market is complete and the no-arbitrage condition holds, the product of the SDF with the value process $I(t)$ of any admissible self-financing trading strategy implemented by trading on a financial asset must be a martingale process,
\begin{equation}
\Lambda(t)I(t)=E[\,\Lambda(t')I(t')\mid \mathscr{F}_t\,] \qquad \forall t'>t~,
\label{htjotjgwg}
\end{equation}
or rewritten in differential form
\begin{equation}\label{E: martingale}
E_{t_0}[\,d(\Lambda(t)I(t))\,]=0~,
\end{equation}
where the expectation operator $E_{t_0}[\,\cdot  \,]$ represents the expectation conditional on all current disclosed information corresponding to the $\sigma$-algebra $\mathscr{F}_{t_0}$. From condition \eqref{E: martingale}, we obtain
\begin{equation}
\begin{aligned}
0 &=E_{t_0}[\,\frac{d(\Lambda_t I_t)}{\Lambda_t I_t}\,]=E_{t_0}[\,\frac{d\Lambda_t}{\Lambda_t}+\frac{dI_t}{I_t}+\frac{d\Lambda_t}{\Lambda_t}\frac{dI_t}{I_t}\,]\\
   &=\{-rdt-\rho_Y E_{t_0}(dY)\}+\{E_{t_0}({\mu}(t))dt+\sigma_Y E_{t_0}(dY)-\kappa h(t)dt\}-\underset{i,j=Y,W}{\sum\sum}\rho_i\sigma_jdt\\
   &=E_{t_0}({\mu}(t))dt-rdt-\kappa h(t)dt-\underset{i,j=Y,W}{\sum\sum}\rho_i\sigma_jdt+(\sigma_Y-\rho_Y)E_{t_0}(dY)\\
   &=E_{t_0}({\mu}(t))dt-rdt-\kappa h(t)dt-\underset{i,j=Y,W}{\sum\sum}\rho_i\sigma_j dt+(\sigma_Y-\rho_Y)(-\alpha e^{-\alpha(t-t_0)}Y_{t_0})dt\,.
\end{aligned}
\end{equation}
The term $\underset{i,j=Y,W}{\sum\sum}\rho_i\sigma_j$ is the required excess return remunerating
all risks at the exception of the crash risk associated with the jump of amplitude $\kappa$. We will
denote it as $\rho\Sigma$ for short. Then, the above equation leads to
\begin{equation}
E_{t_0}({\mu}(t))=(r+\rho\Sigma )+\kappa h(t)+\alpha(\sigma_Y-\rho_Y)e^{-\alpha(t-t_0)}Y_{t_0}~.
\label{E: drift}
\end{equation} 
The dynamics of the crash hazard rate $h(t)$, given by $E_{t_0}[dj]=h(t)dt$, plays a very important here,
as it does in the JLS model. Expression (\ref{E: drift}) includes the expected excess return 
that needs to remunerate rational investors for being exposed to the risk of a crash, which can
occur with the hazard rate $h(t)$. Here as in the JLS model, we assume that the
crash hazard rate $h(t)$ is driven by the behavior of ``noise traders'', who herd into 
successive phases of euphoria and panics. Assuming a dynamics of local imitations and
herding on a hierarchical network of social influences as in the JLS model, this leads
to the crash hazard rate following a LPPL (log-periodic power law) process of the type (\ref{E: landau 1}).

Compared with the JLS model, the new ingredient $Y$ in (\ref{E: Y of OU}) translates
into an additional term proportional to $e^{-\alpha(t-t_0)}Y_{t_0}$ in expression (\ref{E: drift}).
Rather then being deterministic as in the JLS model, the return $E_{t_0}({\mu}(t))$
that is anticipated at time $t_0$ for the time horizon up to $t$ is a function of the specific
stochastic realization $Y_{t_0}$ of the O-U process $Y$ which is known at $t_0$. 
This property captures the possible updates of belief of RE investors. 
Even though RE assumes that a RE investor always makes an unbiased estimation of 
the actual return, it is rational to account for the fact that his/her belief would adjust
to the flow of available information, i.e., ${\mu}(t_1)=E_{t_1}\left( \frac{dI_{t_1}}{I_{t_1}}\right )\ne E_{t_2}\left(\frac{dI_{t_2}}{I_{t_2}}\right)=\mu(t_2)$, for $t_1\ne t_2$.

Since $E_{t_0}(Y_t)= e^{-\alpha(t-t_0)}Y_{t_0}$ by construction of the O-U process $Y_t$, 
the simplest specification for the drift term $\mu(t)$ of the price process (\ref{E: RE model}),
which is compatible with (\ref{E: drift}), reads 
\begin{equation}\label{E: drift 2}
{\mu}(t)=(r+\rho\Sigma)+\kappa h(t)+\alpha(\sigma_Y-\rho_Y )Y_{t} ~.
\end{equation}
Substituting \eqref{E: drift 2} into \eqref{E: RE model}, we obtain
\begin{equation}
\frac{dI}{I}=[r+\rho\Sigma+\kappa h(t)\,]dt-\alpha\rho_Y Y dt+(\sigma_Y +\sigma_W) dW
\label{tkhwpmgfl}
\end{equation}
Similarly, substituting \eqref{E: drift 2} into \eqref{E: RE model 2}, we obtain the discrete formulation
for the dynamics of the logarithmic returns:
\begin{subequations}
\begin{align}
\ln I_{t+1}-\ln I_t & = {\mu}_t + \sigma_Y(Y_{t+1}-Y_t)+\sigma_W \varepsilon_t\notag  \\
                                   & =[r+\rho\Sigma+\kappa h(t)]-\alpha\rho_Y Y_t+(\sigma_Y+\sigma_W)\varepsilon_t 
                                    \label{E: diff I 1}\\
                                   & =[r+\rho\Sigma+\kappa h(t)]+ \rho_Y(Y_{t+1}-Y_{t})+(\sigma_Y+\sigma_W-\rho_Y)\varepsilon_t  \label{E: diff I 2}
\end{align}
\end{subequations}
As explained below equation (\ref{E: drift}), following the JLS model, we assume that
the crash hazard rate $h(t)$ follows a deterministic time-dependence, that describes
the collective behavior of noise traders approaching a critical time at which the
probability per unit time for a crash to occur peaks sharply. Using a model of
social imitation on a hierarchical network of social influences, JLS obtained 
a crash hazard rate obeying a LPPL process. Since $r$, $\rho$, $\Sigma$ and $\kappa$
are assumed constant, the term $r+\rho\Sigma+\kappa h(t)$ is following
a LPPL deterministic process $\Delta H(t)=H(t+1)-H(t)$, where 
$H(t)$ is given by expression (\ref{E: landau 1}).

Then, using  \eqref{E: diff I 2}, the residual $\nu_t \equiv  \ln I_t -H(t)$ of the logarithm of the asset value
with respect to the deterministic LPPL process is given by
\begin{equation}\label{E: diff resi}
\nu_{t+1}-\nu_{t}=\rho_Y(Y_{t+1}-Y_{t})+(\sigma_Y+\sigma_W-\rho_Y)\varepsilon_t~.
\end{equation}
Operationally, the process $\nu_t$ is nothing but the residuals of the nonlinear calibration of 
the process $H(t)$ to the asset price time series $\ln I_t$.

We make the hypothesis that price regimes where bubbles dominate are characterized by
a strong deterministic component $H(t)$ in the log-price dynamics. As a consequence, one
can expect that the residuals $\nu_{t}$ remain bounded, so that the log-price remains
``guided'' by $H(t)$. If $H(t)$ was stochastic, we would say that $\ln I_t$ and $H(t)$
are cointegrated \citep{Granger1991}. Translated in the context of expression (\ref{E: diff resi}),
this implies that we consider the case where $\sigma_Y+\sigma_W  \approx \rho_Y$
with $|\sigma_Y+\sigma_W  - \rho_Y| \ll \rho_Y$. In this limit, the residuals $\nu_t$ are 
stationary and can be taken proportional to $Y_t$, i.e., they follow an 
AR(1) process. Thus, we assume 
\begin{equation}\label{E: diff nu}
\Delta\nu_t=\nu_{t+1}-\nu_t=-\alpha\nu_t+u_t
\end{equation}
where $u_t$ is a Gaussian white noise. From \eqref{E: diff I 2} and \eqref{E: diff resi} and 
using the definition of $\Delta H(t)$, we get
\begin{equation}\label{E: diff I H nu}
\ln I_{t+1}-\ln I_t=\Delta H(t)+\Delta\nu_t~.
\end{equation}
Combining \eqref{E: diff I H nu} and \eqref{E: diff nu}, the recursive formula for the logarithmic asset prices reads 
\begin{equation}\label{E: return adjust}
\ln I_{t+1}=\ln I_{t}+\Delta H_t-\alpha(\ln I_t - H_t)+u_t~.
\end{equation}
Equivalently, the equation for the logarithmic return is
\begin{equation}\label{E: JLSL}
r_{i+1}=\ln I_{t_{i+1}}-\ln I_{t_i}\,\sim\,N(\Delta H_{t_{i+1},t_{i}}-\alpha(\ln I_{t_i} - H_{t_i}),\,\sigma_u^2 (t_{i+1}-t_i)),~~ \Delta H_{t_{i+1},\, t_{i}}=H_{t_{i+1}}-H_{t_{i}}~.
\end{equation}
Compared with the conditional probability distribution given by expression \eqref{E: r_JLS} valid for the JLS model, our model introduces a new stochastic term in the drift. This new term $\alpha(\ln I_{t_i} - H_{t_i})$ ensures that the log-price fluctuates around while remaining in the neighborhood of the LPPL trajectory $H_t$. This formulation ensures the consistency of modeling the log-price by the deterministic LPPL component as a global observable 
emergent macroscopic characteristics. We refer to model (\ref{E: JLSL}) as the ``volatility-confined LPPL model.''
Obviously, this modeling strategy leading to the general form (\ref{E: JLSL})  holds for arbitrary deterministic models $H_t$.

\subsection{Derivation  based on the concept of the {\em Stochastic Discount Factor (SDF) with critical behavior}}

We now present an alternative derivation of the volatility-confined model (\ref{E: JLSL}) with 
a LPPL drift trajectory (\ref{E: landau 1}), from 
a completely different angle compared with the RE bubble model of the previous section. Our 
alternative derivation describes the dynamics of the impact of herding investors 
on asset prices via a novel specification of the stochastic discount factor. This different approach
is motivated by several weaknesses of the RE model.
\begin{itemize}
\item The RE model segments rather artificially the respective roles of noise traders
on the one hand and of RE investors on the other hand. The former are
assumed to control the crash hazard rate only via their herding behavior, 
and their impact on price is indirect through the no-arbitrage condition
representing the actions of RE investors 
that link the conditional expected return to the crash hazard rate.
\item Within the logic of the RE model, notwithstanding the deterministic predictability 
of the crash hazard rate obtained via the corresponding deterministic price component,
the RE investors cannot on average make profit: the RE investors
are remunerated from taking the risk of being exposed to a crash. Over all possible scenarios,
their expected gain is zero. But RE agents endowed with different preferences could
in principle arbitrage the risk-neutral agents. The homogeneity of the RE agent preferences
is therefore a limitation of the model.
\end{itemize}

Rather than using the interplay between the noise traders driving the crash hazard rate
and the risk-adverse rational investors acting as market makers, we 
attribute the characteristics of the price behavior to 
the internal dynamics of the {\em market  sentiment}.  
We propose to capture the critical behavior of an asset price resulting
from the emergent collective organization of the complex financial system
by a specification of the stochastic discount factor (SDF).

The starting point is to recognize the existence of critical dynamics (in the sense of complex
systems) occurring in financial markets. The critical dynamics 
reflects the herding behavior of imitational 
investors, which leads to increasing correlations between the agents translated
into financial bubbles. Such behaviors result from 
imperfective information, the use of heuristics 
and possible biases  in the judgements of heterogeneous investors.
It is therefore natural to combine insights from the field of behavioral finance
and the concepts of criticality developed in the theory of complex systems \citep{Criticalbook}.

From a behavioral finance perspective, we refer to
Shefrin, who extended the SDF into a so-called {\em Behavioral SDF (BSDF)}.
The BSDF is supposed to provide a behaviorally-based synthesis of different theories of asset pricing
\citep{BSDF}. In this approach, the BSDF can be interpreted as
a market sentiment factor, which according to Shefrin, is not a scalar but a stochastic process reflecting the deviation of subjective beliefs described by a certain representative agent (the market itself) relative to objective beliefs and of market's equilibrium time discount factor relative to the situation when all investors hold objectively correct beliefs. Expressed with discrete times, the BSDF can be defined as
\begin{equation}\label{E: BSDF d}
\Lambda^{\mathsf{ST}}(x_t)=\frac{\pi(x_t)}{\Pi(x_t)}=\left[\frac{P_R(x_t)}{\Pi (x_t)}\cdot \frac{\delta_R^t}{\delta_{R,\Pi}^t}\right]\,\cdot \delta_{R,\Pi}^t \,[g(x_t)]^{-\gamma_R(x_t)}~,
\end{equation}
where the exponent $^{\mathsf{ST}}$ stresses that the BSDF embodies the ``sentiment'' of the market.
The term $\pi(x_t)$ denotes the price of a contract that promises a unit-valued payoff, should event $x_t$ occurs at time $t$. $\pi(x_t)$ is thus the state price of the basic security associated with the time-event pair $(t, x_t)$. $\Pi$ denotes the objective probability density and $P_R$ is the representative investor's subjective belief density distribution, which can be derived by aggregating the heterogeneous investor's subjective beliefs given a set of adequate state prices. $\gamma_R$ denotes the coefficient of relative risk aversion of the market. $g$ is the interest rate used to discount future payoffs. The term $\frac{P_R(x_t)}{\Pi (x_t)}\cdot \frac{\delta_R^t}{\delta_{R,\Pi}^t}$ is the product of the deviation of market's subjective beliefs to objective beliefs and of market's equilibrium time discount rate relative to the objective discount rate. Therefore, it plays the role of a  {\em market sentiment} factor, which we denote by $\Phi(x_t)$ below. Notice that the remaining terms of equation \eqref{E: BSDF d} correspond to the traditional SDF, which we still denote by $\Lambda$. This leads to express $\Lambda^{\mathsf{ST}}(x_t)$ as the product of 
$\Phi(x_t)$ and $\Lambda$, or in continuous time, as
\begin{equation}\label{E: BSDF}
\Lambda(t)^{\mathsf{ST}}=\Phi(t)\Lambda(t)~,
\end{equation}
with
\begin{equation}
\Phi(x_t) \equiv \frac{P_R(x_t)}{\Pi (x_t)}\cdot \frac{\delta_R^t}{\delta_{R,\Pi}^t}~,~~~~
\Lambda(t) = \delta_{R,\Pi}^t \,[g(x_t)]^{-\gamma_R(x_t)}~.
\end{equation}

Armed with this representation (\ref{E: BSDF}), we propose to capture the market critical behavior
through the dynamics of the market sentiment factor, which is assumed to be characterized by the following jump process
\begin{equation}\label{E: market st}
\frac{d\Phi_t}{\Phi_t}=a\,dt-b\,dj~.
\end{equation}
The coefficient $a$ is assumed to be small, as it describes the amplitude of the deviations
of the market's equilibrium discount rate from the objective discount rate in ``normal'' times.
In contrast, the term $dj$ governs the occurrence of a possible catastrophe of the market sentiment resulting from a critical collective amplification of pessimism leading to a run-away. When the market operates close to a critical point, increasing crowds of bearish investors gather
in their social imitational network to drive down the market's sentiment which may, as a result, fall
sharply with some probability. For all state $x_t$ except the most extreme jump-crash associated with state $x^{\mathrm{ex}}$, we have $P_R(x_t) <\Pi(x_t)$, i.e., investors underestimate the risks.
On the other hand,  $P_R(x^{\mathrm{ex}})>\Pi(x^{\mathrm{ex}}))$, which means that
the whole market becomes over-pessimistic at the time when the extreme event is revealed.
We also assume that the expectation $E_t(dj)=h(t)dt$ of the jump process $dj$ defines the hazard rate $h(t)$. 
The difference with the RE model of subsection \ref{rederiv} is that, here, 
$h(t)$ represents the probability for an overwhelming synchronized bear raid to occur, conditional on the fact
that the raid has not yet happened. As in subsection \ref{rederiv}, we assume that $h(t)$
follows a deterministic time-dependence with LPPL properties that are typical of a 
critical behavior on a hierarchical network.
Using \eqref{E: BSDF} and \eqref{E: market st}, we have
\begin{equation}\label{E: c BSDF}
\frac{d\Lambda_t^{\mathsf{ST}}}{\Lambda_t^{\mathsf{ST}}}=\frac{d(\Phi_t \Lambda_t)}{\Phi_t \Lambda_t}=\frac{d\Phi_t}{\Phi_t}+\frac{d\Lambda_t}{\Lambda_t}+\frac{d\Phi_t}{\Phi_t}\frac{d\Lambda_t}{\Lambda_t}=-[\,r-a\,]dt-b\,dj+\rho_Y dY+\rho_W dW~.
\end{equation}

For the price process, we use the same model (\ref{E: RE model 2}) as in the previous
subsection and the same process \eqref{E: SDF} for the SDF $\Lambda(t)$.
The main difference with the RE model of subsection  \ref{rederiv}  is that the dynamics
of the asset price given by (\ref{E: RE model 2}) does not have a jump term. 
Since we attribute the possible occurrence of a crash to a phase transition resulting
from a herding behavior, it is in accord with intuition that the inherent process of the asset
price dynamics does not contain jumps.

Assuming that the financial market is complete and in absence of risk-free arbitrage, the
product of the rate of change of asset price and of the BSDF should satisfy the martingale ccondition, i.e., $E_t[d(\Lambda_t^{\mathsf{ST}}I_t)]=0$. With \eqref{E: c BSDF} and \eqref{E: RE model 2}, this leads to 
\begin{equation}
\frac{dI}{I}=[r+\rho\Sigma-a+b\, h(t)\,]dt-\alpha\rho_Y Y dt+(\sigma_Y+\sigma_W )dW~.
\label{tkhwpmgfl_2}
\end{equation}
This equation has the same structure as expression (\ref{tkhwpmgfl}) obtained with the RE model,
with just a redefinition of the constants $r+\rho\Sigma \to r+\rho\Sigma-a$ and $\kappa \to b$.
With the same price dynamics, the conditional probability distribution of returns are identical. 
It is this model (\ref{tkhwpmgfl_2}) or equivalently  (\ref{tkhwpmgfl}) that we will calibrate
and test in the next sections.

But, before doing so, let us interpret the economic meaning of the above 
derivation based on the concept of the BSDF with critical behavior.
In contrast with the RE model, there is not need here for a representative rational investor
playing the role of a market maker fixing the price on the basis of his rational expectations. The
underlying origin of the martingale condition and the mechanism for the crash are quite different from that of the RE model. First, we assume that the financial market is complete, i.e., {\em Arrow-Debreu securities (A-Ds)} are available to all investors that allow a perfect replication of the asset value before the crash.
In the early stage of  the bubble regime, as the whole market is over-optimistic, the probability for a sharp
price drop is underestimated and taken to be vanishingly small. Hence, the price of A-Ds for a crash state is also zero. In this situation, the current price of the asset is the aggregation of the prices for all available A-Ds that correspond to all expectable states of price variations in the market. As time goes on, the percentage of bearish investors becomes larger and larger, as the deviation of the asset price from its fundamental value increases. When the fraction of bearish investors approaches the critical value from below, with some non-zero probability, the market sentiment may shift to over-pessimic and, as a consequence, trigger a sudden jump. This jump occurs as a result of 
amplified subjectively perceived probability for a crash, embodying the now predominant over-pessimistic bias.
Because there are not yet A-Ds associated to the states corresponding to very sharply declining prices, 
nobody is able to hedge this extreme risk. Therefore, there is a tension hovering over the market, which is
described by the hazard rate $h(t) \equiv E_t(dj)/dt$, where $dj$ punctuates the dynamics 
(\ref{E: market st}) of the {\em market sentiment} factor. The existence of the hazard rate leads
investors to require higher returns to compensate for their risks\footnote{In this model, the stock price in the bubble regime is risk driven. But quite different from the RE model in which only the representative RE investor requires a compensation for his exposition to market risks, here all investors in the market, irrespective of whether they are rational or irrational, are collectively requiring
higher and higher returns as the bubble develops.}. This is implemented by the
martingale condition, expressing that there is no opportunity for  riskless arbitrage. However,  when the 
downward jump happens, all investors suddenly find that the available A-Ds that replicate the asset price have become cheap. Then, it is rational for them to short sell their stock and buy all the A-Ds. Given the absence of A-Ds for extreme drops of stock price, this then leads to an arbitrage opportunity. This results in further price fall, fueled
by the positive feedback of the strategic allocation used by investors (short the asset and long the A-Ds).
The crash is thus the result of the cumulative effect of this vicious circle, corresponding to a spontaneous breaking of equilibrium \citep{Didier2000b}.

\section{Tests based on the Ornstein-Uhlenbeck structure of Residuals of the LPPL model \label{thbogda}}

We now describe a first series of empirical tests performed using model (\ref{tkhwpmgfl_2}) (or equivalently  (\ref{tkhwpmgfl})), supplemented by the LPPL specification (\ref{E: landau 1}).
One key feature is the Ornstein-Uhlenbeck (O-U) structure of the residuals. This suggests that 
evaluations of our model of a bubble regime should test both for the presence of 
significant LPPL signatures as well as for the O-U property of residuals. 
According to \eqref{E: diff nu}, this translates into an AR(1) test for the residuals 
obtained by fitting the asset price trajectory using a LPPL process (\ref{E: landau 1}).
We will therefore use two strategies. The first one developed in this section calibrates the asset price
and then tests for the O-U properties for the residuals. The second one, which is implemented
in section \ref{section4}, uses the equivalent specification 
(\ref{E: JLSL}) on the asset returns to develop a Bayesian inference test.

\subsection{Evaluation of GARCH processes to test for errors of type I (false positive)}

Recall that the purpose of this paper is to test the claim that financial bubbles
can be diagnosed from their super-exponential price dynamics, possibly decorated by
log-periodic accelerating oscillations. A first approach is to test whether standard financial
processes exhibit such signatures. As an illustration, let us consider the GARCH (1,1) 
model
\begin{equation}
\begin{aligned}
\ln I_t-\ln I_{t-1} &=\mu_0+\sigma_t z_t\\
\sigma_t ^2 &=\sigma_0^2+\alpha(\ln I_{t-1}-\ln I_{t-2}-\mu_0)^2+\beta\sigma_{t-1}^2~,
\end{aligned}
\label{gthogrbrgbwq}
\end{equation}
where the innovation $z$ is distributed according to the Student-$n$ distribution (with $n$ degrees of freedom). Estimating this GARCH(1,1) model on the S\&P500 index for the US market  from Jan.~3, 1950 to Nov.~21, 2008
at the daily time scale (such that one unit time increment in (\ref{gthogrbrgbwq}) corresponds to one day)
yields the following  parameters: conditional mean of return $\mu_0=5.4\times10^{-4}$, conditional variance $\sigma_0=5.1\times10^{-7}$, ARCH coefficient $\alpha=0.07$, GARCH coefficient $\beta=0.926$ and number of degrees of freedom of the student distribution close $n=7$. 

Calibrating the LPPL specification (\ref{E: landau 1}) to a given price trajectory will always provide some output
for the parameters and the residuals. In order to qualify the LPPL calibration, we impose the following
restrictions on the parameters
\begin{equation}\label{E: LPPL condition}
\begin{aligned}
B &>0\\
0.1\le\beta &\le 0.9\\
6\le\omega&\le13\\
|C|&<1\end{aligned}
\end{equation}
These conditions (\ref{E: LPPL condition}) can be regarded as the ``stylized features of LPPL'', 
which were documented in many previous investigations (see \citet{Johansen2004} and \citet{Didier2006}
for reviews documenting these stylized facts). The two first conditions 
$B >0$ and $0.1\le\beta \le 0.9$ ensures a faster-than-exponential acceleration of the log-price
with a vertical slope at the critical time $t_c$. The condition $6 \le\omega \le13$ constrains the
log-periodic oscillations to be neither too fast (otherwise they would fit the random component
of the data), nor too slow (otherwise they would provide a contribution to the trend, see \citet{Huang2000}
for the conditions on the statistical significance of log-periodicity). The last restriction $|C|<1$ in \eqref{E: LPPL condition} was introduced by \citet{Bothmer} to ensure that the hazard rate $h(t)$ 
remains always positive. For the sake of brevity,  we refer to conditions (\ref{E: LPPL condition}) 
as the {\em LPPL conditions}.  We also impose the search of the critical time $t_c$ to be
no further than one year beyond the last data point used in the fit.

Table \ref{T: 1} shows the results obtained by calibrating the  LPPL specification (\ref{E: landau 1})
to synthetic time series generated with the GARCH model (\ref{gthogrbrgbwq}), with the 
LPPL conditions (\ref{E: LPPL condition}), and the unit-root tests on the residuals. We have
performed these tests on two sets of 1000 synthetic GARCH time series: (i) samples of random lengths,
with lengths uniformly distributed from $750$ days to $1500$ days and (ii) samples of fixed length of $1500$ days. 
The unit-root tests are the Phillips-Perron test and the Dickey-Fuller test, which are such that a rejection of the 
null hypothesis $H_0$ implies that the residuals are stationary (and therefore are compatible with the Ornstein-Uhlenbeck process posited in our model presented in the previous section \ref{s:2}). Table \ref{T: 1} shows first a very small rate of false positives, i.e., less than $0.2\%$ of the 2000 GARCH-generated time series are found to obey
the LPPL conditions, and would thus be diagnosed as being in a bubble regime. Secondly, the unit-root tests
show that, for most residual time series obtained as the difference between the synthetic GARCH time series and their LPPL calibration, one can not reject the null, i.e. the residuals are non-stationary. This confirms that our model
is not a good fit to synthetic GARCH time series.

\subsection{Tests on the S\&P500 US index from Jan.~3, 1950 to Nov.~21, 2008}

We now apply the same procedure as in the previous subsection to 
the S\&P500 index in the US from Jan.~3, 1950 to Nov.~21, 2008. But we do not
have of course the luxury of a large sample of different realizations, as for the synthetically
generated GARCH time series. Instead, we generate two sets of time windows of $750$ successive
trading days. The first (respectively second) set is obtained by sliding windows of $750$ days
over the whole duration of our data sets with time increments of 
25 days (respectively 50 days), referred to as windows of type I and II respectively. The
first (second) set has 563 (262) windows. 

In table \ref{T: 2}, we can see that, for set I (respectively II), a fraction $P_{\mathrm{LPPL}}=2.49\%$ (respectively $2.84\%$) of the windows obey the LPPL conditions (\ref{E: LPPL condition}). This is more than a factor of ten larger than the corresponding fraction for the synthetic GARCH time series. For this fraction of windows which
obey the LPPL conditions, all of them reject the two unit-root tests for non-stationarity, showing that the
time windows, that qualify as being in a bubble regime according to our model, also give residuals which
are stationary, as required from the Ornstein-Uhlenbeck specification of the residues of our model.
In contrast, table \ref{T: 2} shows that, as for table \ref{T: 1}, the large majority of windows
give residuals for which the null unit-root hypothesis of non-stationarity cannot be rejected. This means
that, for most windows that do not obey the LPPL conditions, their residuals are non-stationary, 
providing two reasons for diagnosing these windows as being in a non-bubble regime. This result,
together with the 100\% rate of rejection of the null hypothesis for non-stationarity for the 
subset of windows which obey the LPPL conditions, provides a strong support for our model.
In contrast, for windows that are diagnosed to be in a bubble regime, their residues are
automatically stationary, in accordance with our model. A crucial additional evidence
is provided by table \ref{T: 3} which lists
the windows that obey the LPPL conditions. We find that all of them correspond to 
periods preceding well-known crashes. This confirms that our method
for identifying bubbles exhibits a very low rate of errors of type I (false positives).

Summarizing our results obtained so far, we can state that about 97-97.5\% of the time intervals
of 750 trading days within the period from Jan.~3, 1950 to Nov.~21, 2008 correspond
to non-bubble regimes, rather well described by a GARCH process. We have been able
to characterize LPPL signatures of bubbles that occupy about 2.5-3\% of the whole time interval.
These percentages suggest a highly selective and efficient detection filter. 
We test further this selectivity by focusing on the classic crash of October 1987, to test
how well we can diagnose a bubble regime preceding it.
We consider shrinking windows with increasing starting dates and fixed last date of Sep. 30 1987. 
We scan the starting dates with a resolution of 5 days and stop with the shortest window
of size equal to 750 trading days. We expect that the LPPL conditions 
and the rejection of the null unit-test hypothesis for the residuals should be observed
increasingly as the starting date of the windows moves upward towards the crash date.
Table \ref{T: 4} shows the results for different starting dates, which confirm remarkably
well our expectations. The closer the starting date is to the crash date, the larger is
the fraction $P_{\mathrm{LPPL}}$ of windows that obey the LPPL conditions. Of these,
a fraction of $P_{\mathrm{Stationary Resi.}\mid \mathrm{LPPL}}=100\%$ 
reject the null unit-root tests of non-stationarity. Compared with the overall fraction
of $2.5-3\%$ of windows that pass the LPPL conditions over the whole
time interval from Jan.~3, 1950 to Nov.~21, 2008, this fraction rises drastically from  about 20\% to 100\%
for the time windows most influenced by the latest part of the time series closest to the crash. 
This suggests the existence of a regime shift from a GARCH-like process to a LPPL bubble regime
as time approaches the Oct. 1987 crash. Note also that all 43 windows that pass the LPPL conditions
have starting dates around the end of 1983, suggesting that the bubble that led to the great Oct. 1987 crash
started around the beginning of 1984. This results is very interesting in so far that it 
strengthens the interpretation of crashes as the outcome of a long maturation 
process, and not due to proximal causes of the previous few days or weeks.

\begin{figure}[!h]
\begin{center}
\includegraphics[width=16cm]{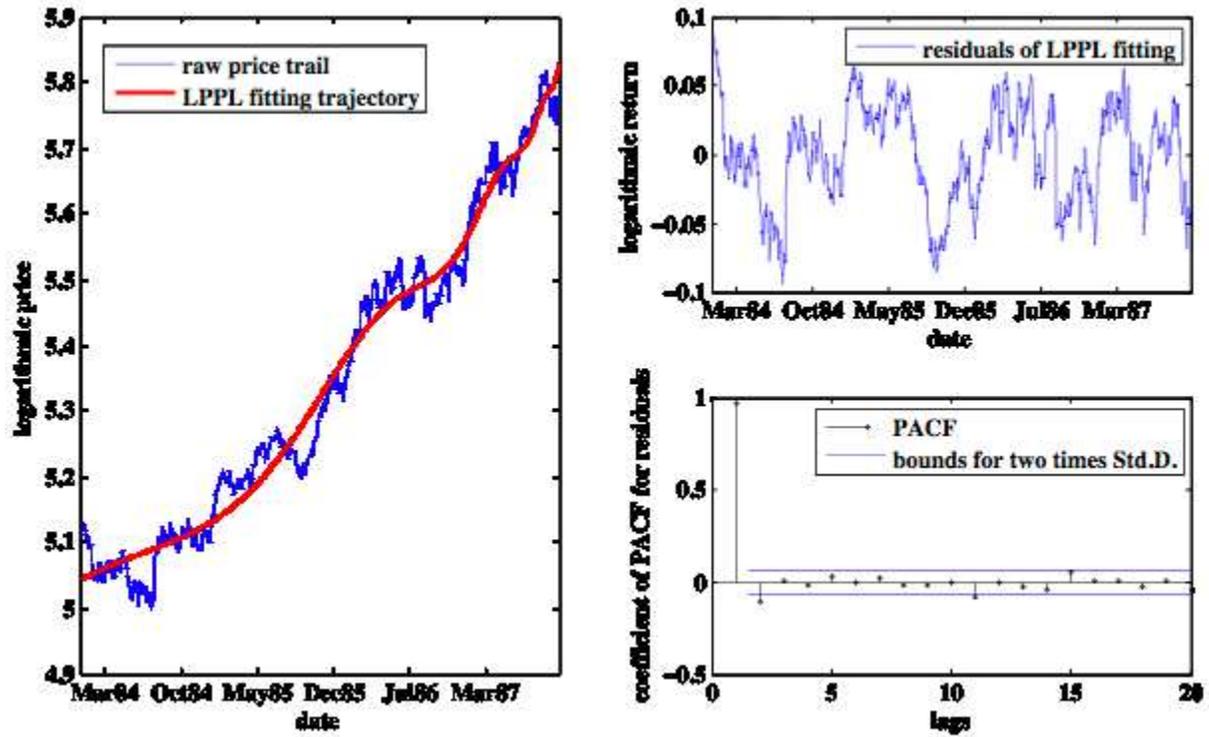}
\caption{Left panel:  fit of the logarithm of the S\&P500 US index with 
expression (\ref{E: landau 1}) over the time interval from Jan.~3,  1984 (the first trading day in 1984) to Sep.~30, 1987.
Upper right panel:  time series of the residuals of the fit shown in the left panel. Lower right panel: partial autoregression correlation function (PACF) of the residuals. The value of the PACF at lag 1 is equal to 0.9709. For lags larger than 1, the PACF is bounded between $\pm$ two standard deviations. }
\label{F: 1}
\end{center}
\end{figure}

The left panel of Fig.~\ref{F: 1} shows the fit of the logarithm of the S\&P500 US index with 
expression (\ref{E: landau 1}) over the time interval from Jan.~3,  1984 (the first trading day in 1984) to Sep.~30, 1987.
The time series of the residuals of this fit is shown in the upper right panel and its 
partial autoregression correlation function (PACF) is depicted in the lower right panel for lags from 0 to 20 days.
All values of the PACF with lags larger than $1$ fall within two standard deviations, indicating the 
absence of linear dependence. Combined with 
the result of the Phillips-Perron test on this series of residuals shown in Table \ref{T: 5}, this suggests 
that these residuals are both stationary (they reject the unit root test of non-stationarity) and furthermore they
can be closely approximated by an AR(1) process with a mean-reverting coefficient $-\alpha \approx -0.03$.
This supports our proposal to model the residuals $\nu(t)$ of the LPPL as generated by a Ornstein-Uhlenbeck process.

\section{Bayesian inference for our modified LPPL model with Ornstein-Uhlenbeck residuals  \label{section4}}

We now describe the second series of empirical tests performed using model (\ref{tkhwpmgfl_2}) (or equivalently  (\ref{tkhwpmgfl})), supplemented by the LPPL specification (\ref{E: landau 1}).
While the previous section \ref{thbogda} has used the asset price to test for the presence
of LPPL conditions and has then tested for the Orstein-Uhlenbeck (O-U) properties for the residuals, here 
we use the other equivalent specification 
(\ref{E: JLSL}) on the asset returns to develop a Bayesian inference test.

Our approach parallels that of Chang and Feigenbaum (2006) for the implementation of the Bayesian inference. But a fundamental difference is that, while
their implementation used the specification \eqref{E: r_JLS}, our model (\ref{E: JLSL}) contains the additional term $-\alpha (\ln p_{t }-H_{t})$ stemming from the intrinsic guiding mechanism associated with the O-U model of the residuals decorating the deterministic LPPL bubble trajectory. We show below that this new term
makes all the difference in establishing the statistical significance of LPPL properties of asset returns.

\begin{figure}[htdp]
\begin{center}
\includegraphics[width=13cm]{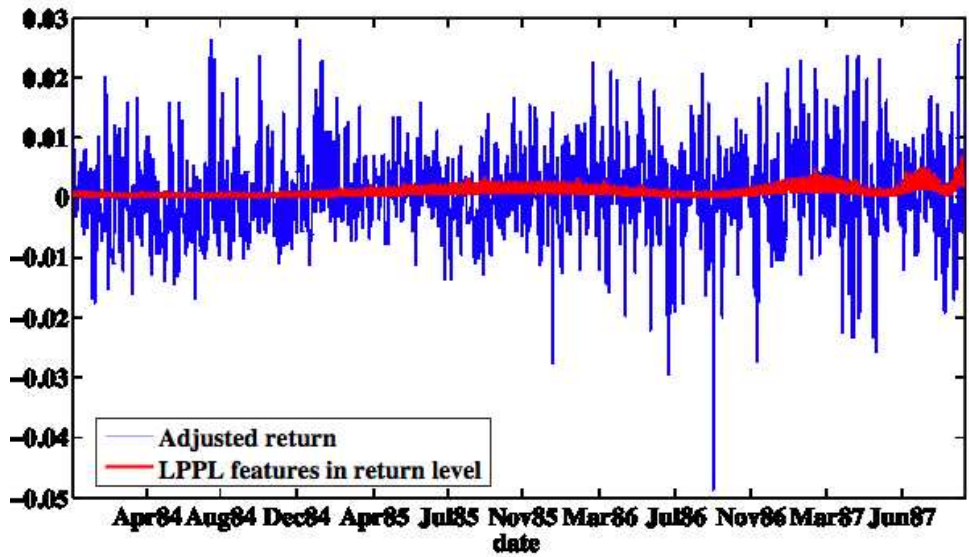}
\caption{Time series of adjusted returns defined by expression (\ref{hbrgonwga})
for the S\&P500 US index from Jan.~3, 1984 to Sept.~30, 1987. The smooth
continuous line shows the LPPL term $\Delta H_{t}$, where
$H_{t}$ is defined by equation (\ref{E: landau 1}). }
\label{F: 2}
\end{center}
\end{figure}

Equation (\ref{E: r_JLS}) suggests that one might detect directly the LPPL signature in returns by removing the effects 
caused by the intrinsic guiding mechanism associated with the O-U model of the residuals.
Defining the random variable $\Psi_{t_i}=-\alpha (\ln p_{t }-H_{t})$, we define the {\em adjusted return} as
\begin{equation}
r_{t_i}^{\mathrm{Ad}}=r_{t}-\Psi_{t}=\Delta H_{t}+u_t~.
\label{hbrgonwga}
\end{equation}
Recall that $\Delta H_{t}$ results directly from the hazard rate and contains the LPPL signal. The
residual $u_t$ should then be a white noise process. The adjusted returns
$r_{t}^{\mathrm{Ad}}$ defined by (\ref{hbrgonwga}) for the 
 S\&P500 US index from Jan.~3, 1984 to Sept.~30, 1987 are shown in Fig.~\ref{F: 2}.
 The continuous curve shows $\Delta H_{t}$, where the parameters for the process 
$H_{t}$ are obtained by a nonlinear least square fit as in the previous section. 
Unsurprisingly, one can 
see that the deterministic component is very small compared with the typical amplitude
of the adjusted returns. Note that the same relative smallness of the LPPL signal 
viewed in the return time series has been noted earlier \citep{Feigenbaum2001, Feigenbaum2006}.
It is not clear how to develop a test that directly test for the existence of a significant 
LPPL component in the time series of adjusted returns shown in Fig.~\ref{F: 2}.

The general weakness
of the likelihood analysis of log-periodicity on returns is not a surprise when viewed from the
perspective offered by the analysis of \citet{Huang2000}. Using numerically intensive Monte-
Carlo simulations, \citet{Huang2000} showed that, for regularly sampled time series as is the
case for financial time series, the log-periodic signal is much more significant in the cumulative
signal than in its first difference (and that using the cumulative signal does not create spurious
log-periodicity), due to the well-known fact that integration corresponds to low-pass filtering.
This suggests that working on returns, while being the standard of econometric studies, may
actually be sub-optimal in this case. \citet{Didier2001b} summarized in their section 9 the Monte-Carlo tests which have been
performed by various groups to address specifically this problem, including \citep{Feigenbaum1996, Feigenbaum1998}, both on synthetically generated price levels and on randomly chosen time
intervals of real financial time series: these tests show the high statistical significance of logperiodicity
in the log-price trajectory before the crash of October 1987 and on several other
bubbles.

We thereupon turn to the method of Bayesian inference to investigate the statistical significance of LPPL features in the return time series. Following the philosophy attached to Bayesian analysis, two models can be compared 
by estimating the ratio of the posterior probability for each model given the data, this ratio being called the {\em Bayesian factor}. Let $M_0$ denotes the benchmark model and $\Xi_0$ its corresponding set of parameters. Similarly, let $M_1$ denotes an alternative model with its set of parameters $\Xi_1$. Then, the Bayesian factor of model $M_1$ compared with model $M_0$ is defined as
\begin{equation}
\begin{aligned}
B_ {M_1, M_0} &=\frac{p(\,\Xi_1 \mid Q \;; M_1)}{p(\,\Xi_0 \mid Q \;; M_0)
}\\
&=\frac{\frac{\int p(\theta_{M_1} ; M_1) p(Q\mid \theta_{M_1} ; M_1) d \theta_{M_1}}{p(Q)}}{\frac{\int p(\theta_{M_0} ; M_0) p(Q\mid \theta_{M_0} ; M_0) d \theta_{M_0}}{p(Q)}}
&=\frac{\int p(\theta_{M_1} ; M_1) p(Q\mid \theta_{M_1}; M_1) d \theta_{M_1}}{\int p(\theta_{M_0} ; M_0) p(Q\mid \theta_{M_0} ; M_0) d \theta_{M_0}}~.
\end{aligned}
\end{equation}%
In this expression, $\theta_M$ denotes the vector of parameters for model  $M$.  The term $p(\,\Xi \mid Q \;; M)$ represents the posterior probability for the set of parameters in model $M$, given the observed data $Q$. The term $p(\theta_{M} ; M)$ is the prior probability chosen for the parameters $\theta$ in model $M$.  
Within the framework of Bayesian hypothesis testing,  if $B_ {M_1, M_0}$ is larger than $1$, one should accept the alternative model because the posterior probability for its parameters has enjoyed a larger increase from its initial prior basis level, which implies that the alternative model can explain the data better than the reference model.
If the prior probabilities are not too restrictive, and for a large sufficient data set, Bayesian inference amounts to comparing the likelihood function of each model and the Bayesian factor test tends asymptotically for large data sets
to the likelihood ratio test.

Let us consider the time series of returns $\{q_i\}$ sampled at the time instants 
$t \in \{t_0, t_1,t_2, \cdots, t_N\}$.  For the reference model, as in \citet{Feigenbaum2006},
we choose the Black-Scholes model whose logarithmic returns are given by
\begin{equation}
r_{i}\,\sim\, N(\mu (t_{i}-t_{i-1}) , \sigma^2(t_{i}-t_{i-1}))~.
\end{equation}
The drift $\mu$ is drawn from the prior distribution $N(\mu_{r}, \sigma_{r})$. The variance $\sigma^2$ of daily returns is specified in terms of its inverse $\tau=\frac{1}{\sigma^2}$, known as the ``precision'' in the language of Bayesian analysis. The precision describes how precisely the random variable will be known and thus the higher the better. The precision is supposed to be drawn as $\tau\sim\,\Gamma(\alpha_\tau, \beta_\tau)$.

The alternate hypothesis model is our volatility-confined LPPL model. Recalling expression \eqref{E: JLSL}
with our present notations, the returns of the alternative model are described by
\begin{equation}\label{E: JLSL 2}
r_{i}\,\sim\,N(\Delta H_{{i},\,{i-1}}-\alpha(q_{\,i-1} - H_{{i-1}}),\,\sigma_u^2 (t_{i}-t_{i-1}))
\end{equation}
where 
\begin{multline}
\Delta H_{{i},\,{i-1}}=B(t_c-t_{i-1})^{\beta}\left[1+\frac{C}{\sqrt{1+(\frac{\omega}{\beta})^2}}\cos(\omega\ln (t_c-t_{i-1}))+\phi)\right]\\-B(t_c-t_{i})^{\beta}\left[1+\frac{C}{\sqrt{1+(\frac{\omega}{\beta})^2}}\cos(\omega\ln (t_c-t_{i}))+\phi)\right]\notag
\end{multline}
The LPPL characteristics of the model for $\Delta H_{{i},\,{i-1}}$ are encoded in the 
the vector of parameters  $\boldsymbol{\xi}=( A,B,C, \beta,\omega,\phi, t_c )$. For 
simplicity, we assume that these parameters are drawn independently from the following prior distributions:
\begin{equation}
\begin{aligned}
A\, &\sim\, N(\mu_A\,,\sigma_A)\\
B\, &\sim\, \varGamma\,(\alpha_B\,,\beta_B)\\
C\, &\sim\, U(0,1)\\
\beta\, &\sim\, B(\alpha_\beta\,\,,\beta_\beta)\\
\omega\, &\sim\, \varGamma(\alpha_\omega\,\,,\beta_\omega)\\
\phi \, &\sim\, U(0\,, 2\pi)\\
t_c-t_N\, &\sim\, \varGamma(\alpha_{t_c}\,,\beta_{t_c})\\
\end{aligned}
\end{equation}
where $\varGamma$, $B$ and $U$ denote the $\varGamma$-distribution, $B$-distribution and uniform distribution respectively. In practice of bayesian inference, the $\varGamma$-distribution and $B$-distribution are often adopted as prior probability distribution\footnote{ $\varGamma$-distribution and $B$-distribution are also called {\em conjugate prior family}, because  by 
adopting a prior density of Beta (Gamma) form one also obtain a posterior density of Beta (Gamma) form, but with different parameters. Although there is no necessity to adopt conjugate prior, the conjugate prior property is very convenient for it avoids having to integrate numerically to find the normalising 
constant in the posterior density \citep{young2005}.}. The $\varGamma$-distribution is usually used to describe non-negative variable, and has the density function is $f(x ; \alpha, \beta)=\beta^{-\alpha}\Gamma^{-1}(\alpha)\,x^{\alpha-1}\exp\left(\frac{x}{\beta}\right)$ , with $E(X)=\alpha \beta$ and $Var(X)=\alpha \beta^2$. $\Gamma(z)$ is the gamma function defined as $\Gamma(z)=\int_0^{\infty}t^{z-1}e^{-t}dt$. The random variable realized between 0 and 1 is usually assigned with beta prior density, which is $f(x; \alpha, \beta)=\frac{1}{B(\alpha,\beta)}x^{\alpha-1}(1-\beta)^{\beta-1}$, where $B(z)$ is beta function satisfying $B(u,v)=\frac{\Gamma(u)\Gamma(v)}{\Gamma(u+v)}$. Accordingly, the mean and variance of the variable with $B$-distribution are $E(X)=\frac{\alpha}{\alpha+\beta}$ and $Var(X)=\frac{\alpha \beta}{(\alpha+\beta+1)(\alpha+\beta)^2}$.
Then, the full set of parameters of the volatility-confined LPPL model is $\Xi=(\mu, \tau, \alpha, \boldsymbol{\xi})$. The prior density for our model is given explicitly by the product of all marginal priors for the each parameter
\begin{multline}
p(\theta_{LPPL}\,; \mathrm{LPPL} )=\frac{1}{\sqrt{2 \pi}\sigma_r}\exp\left[-\frac{(\mu-\mu_r)^2}{2 \sigma_r^2}\right]\times f_{\Gamma}(\tau; \alpha_{\tau}, \beta{\tau})\\\times  f_{\Gamma}(\alpha; \alpha_{\alpha}, \beta_{\alpha})\times \frac{1}{\sqrt{2 \pi}\sigma_A}\exp\left[-\frac{(\mu-\mu_A)^2}{2 \sigma_A^2}\right] \times  f_{\Gamma}(B; \alpha_{B}, \beta_{B})\\\times 
 f_{B}(\beta; \alpha_{\beta}, \beta_{\beta})\times  f_{\Gamma}(\omega; \alpha_{\omega}, \beta{\omega})
\times \frac{1}{2\pi}\times  f_{\Gamma}(t_c-t_N; \alpha_{t_c-t_N}, \beta_{t_c-t_N})~,
\end{multline}
for $\theta_{LPPL}\in \Xi=\mathbb{R}^2\times \mathbb{R}_{+}^{3}\times [0,1]^3\times [0, 2\pi)\times[t_N, \infty)$.
According to \eqref{E: JLSL 2}, given $\theta_{LPPL}$ and $q_{i-1}$, the updated posterior density for $q_{i}$ is
\begin{equation}
p(q_{i}\mid q_{i-1},\theta_{LPPL}\,; \mathrm{LPPL})=\sqrt{\frac{\tau}{2\pi (t_{i}-t_{i-1})}}\exp\left[-\frac{\tau(q_{i}-q_{i-1}+\alpha(q_{i-1}-H_{i-1})-\Delta H_{i,i-1})^2}{2(t_{i}-t_{i-1})}\right]~.
\end{equation}
Thus, the conditional density of the returns given the prior parameters reads
\begin{equation}
 p(Q\mid \theta_{LPPL} ; \mathrm{LPPL})=\prod_{i=1}^{N}p(q_{i}\mid q_{i-1},\theta_{LPPL}\,; \mathrm{LPPL})~,
 \end{equation}
and the log marginal likelihood needed for the computation of the Bayesian factor is given by
\begin{equation}\label{E: int}
\mathscr{L}=\ln \left(\int_{\Xi} p(\theta_{LPPL})p(Q\mid \theta_{LPPL} \,; \mathrm{LPPL})d\theta_{LPPL}\right)~.
\end{equation}
Expression (\ref{E: int}) defines nothing but a smoothing of the likelihood function performed
with respect to some a priori weight for the input parameters.

Before proceeding to the calculation of expression (\ref{E: int}) for 
$\mathscr{L}_{\small{\mathrm{LPPL}}}$ and  $\mathscr{L}_{\small{\mathrm{BS}}}$
and obtain the Bayesian factor, we should point out that a major difficulty with 
the Bayesian inference test lies with the fact that the prior distribution is in general unknown to us.
This difficulty  cannot really be
alleviated by trying different priors and by checking the corresponding posteriors, because all
posteriors are false as long as we do know the true a priori distribution of the parameters. 
We stress that there is a
highly non-trivial assumption underlying the Bayesian inference test, namely that the parameters can be considered as random values: random parameters would need in general an ensemble of different sample
realizations (or series of experiments), whereas we are interested here in one particular
realization (or sample). In a sense, the Bayes approach to hypothesis testing assumes that
some kind of ergodicity on a single sample applies and that the sample is of sufficiently
large size. But this needs to be tested and it is not a trivial task.

Given this, we nevertheless pursue, if only for the goal of comparing with the negative results
of the same procedure applied to the JLS model by \citet{Feigenbaum2006}.
To implement the Bayesian inference test, we consider the same data set as before, namely the S\&P 500 US index , but concentrating on the period from Jan.~3, 1984 to Sep.~30, 1987 to correspond with our previous analysis. The constant drift $\mu$, the precision $\tau$, coefficient B and C, super-exponential $\beta$, circular frequency for log-periodic oscillation $\omega$ and phase term $\phi$  are assigned with the same priors as those in \citet{Feigenbaum2006}.  The coefficient $A$, which is the final
expected price at critical time, is taken from a normal distribution with $E[A]=6$ and $Var[A]=0,05$ to roughly accord with the extend of price fluctuations near the critical time. Since $t_c$ can be a few days or months after the real crash, but with the most probable  value just being the crash day, we choose $E[t_c-t_N]=30$ and standard deviation as $\sqrt{Var[t_c-t_N]}=30$. Additionally, we choose $E[\alpha]=\sqrt{Var[\alpha]}=0.05$, which roughly reflects our estimated results obtained from the test using shrinking windows with a fixed last date of Sep.~30 1987 and with time increments of 5 days. The following gives the priors:
 \[
\begin{aligned}
\mu \, &\sim\, N(0.0003, (0.01)^2)\\
\tau \, &\sim\, \varGamma\,(1.0\,,10^{5})\\
\alpha\, &\sim\, \varGamma\,(1.0\,,0.05)\\
A\, &\sim\, N(6\,,0.05)\\
B\, &\sim\, \varGamma\,(1\,,0.01)\\
C\, &\sim\, U(0,1)\\
\beta\, &\sim\, B(40\,\,,30)\\
\omega\, &\sim\, \varGamma(16\,\,,0.4)\\
\phi \, &\sim\, U(0\,, 2\pi)\\
t_c-t_N\, &\sim\, \varGamma(1\,,30)\\
\end{aligned}
\]

The integrals in \eqref{E: int} for the log marginal likelihood 
have been estimated by the Monte-Carlo method with 
10000 sampling values for each integral component. In order to ascertain the validity of our
numerical estimation of $\mathscr{L}_{\small{\mathrm{LPPL}}}$ in \eqref{E: int} 
and to estimate its confidence interval, we
have repeated these calculation 100 times. We also performed the same calculations for
$\mathscr{L}_{\small{\mathrm{BS}}}$  and finally get
\begin{equation}
\begin{aligned}
\mathscr{L}_{\small{\mathrm{LPPL}}}(2.5\%-97.5\%) &=3173.546  - 3176.983\\
\mathscr{L}_{\small{\mathrm{BS}}}(2.5\%-97.5\%) &=3169.808 - 3170.097~.
\end{aligned}
\end{equation}
A difference of the average loglikelihood $\bar{\mathscr{L}}_{\small{\mathrm{LPPL}}} - \bar{\mathscr{L}}_{\small{\mathrm{BS}}}$ of
about $5$ translates into a very large Bayesian factor $\exp(\bar{\mathscr{L}}_{\small{\mathrm{LPPL}}} - \bar{\mathscr{L}}_{\small{\mathrm{BS}}} ) \approx e^5 \simeq 150$. The Bayesian inference test
therefore suggests that our volatility-confined LPPL model strongly outperforms the Black-Scholes benchmark.

Our result contrasts decisively with that of \citet{Feigenbaum2006}. Using our
numerical scheme, we were able to reproduce the negative results reported by \citet{Feigenbaum2006} that the JLS model is not significantly preferred to the benchmark model according to the
Bayesian inference test. Thus, our new results cannot be ascribed to a spurious 
numerical implementation but reveals the importance of the specification of the residuals.
The difference can be 
traced back to the Ornstein-Uhlenbeck model of residuals, which make the LPPL fits self-consistent.
Given the empirical price data, any agnostic economist would have to put more weight on our volatility-confined LPPL model than on the standard benchmark without super-exponential growth and log-periodicity. 

In addition, we calculate the log marginal likelihood for the volatility-confined PL (power law) model.
The PL model is the special case of the volatility-confined LPPL model obtained for $C=0$ in expression \eqref{E: landau 1}. The PL model keeps the super-exponential component but neglects
the log-periodic oscillatory component. The following compares the
log-likelihoods of the two models in their 2.5-percentile to 97.5-percentile range obtained
over the distribution of their numerical estimations:
\begin{equation}
\begin{aligned}
\mathscr{L}_{\small{\mathrm{LPPL}}}(2.5\%-97.5\%) &=3173.546  - 3176.983\\
\mathscr{L}_{\small{\mathrm{PL}}}(2.5\%-97.5\%) &=3175.520 - 3178.425~.
\end{aligned}
\end{equation}
This shows that there is no significant gain 
in the Bayesian factor when going from the PL model to the LPPL model,
Actually, the Bayesian factor for the volatility-confined PL  model tends to be somewhat larger
than that of the volatility-confined LPPL model. This should probably be attributed to the stronger impact
of the priors of the later due to its larger number of parameters, compared with the former.

Indeed, since the Bayes approach suggests to
smooth out the likelihoods corresponding to different parameter values by an a priori density, it
is a legitimate question to ask why such smoothing may work. When the sample size $n$ tends to
infinity, the maximum-likelihood ML-estimates tend to the true values and the likelihood function under the integral
in (\ref{E: int}) ``cuts out'' only a narrow neighborhood of the true values. Thus, the behavior of the a
priori density outside of this neighborhood becomes irrelevant, and the Bayes approach tends to
the maximum likelihood approach, of course under the condition that the chosen prior would not
ascribe zero weight to the true parameter value. However, when the sample size is moderate or
small and the number of parameters is not small, the situation becomes more and more uncertain. 
The likelihoods can have several
(even many) local maxima in the present case of log-periodicity. Proponents of the Bayes
approach argue that this multiplicity is overcome by integration (smoothing). But, for finite
sample size $n$, the smoothing in the marginal likelihood may be more harmful (in particular
under unfortunate choices of the prior): smoothing and its positive effects (suppression or
decreasing multiplicity of local peaks) come at the price of a loss of efficiency. We believe
this could explain the somewhat better performance of the LP model compared with the LPPL model
within the Bayesian inference tests.

In conclusion, we find a decisive preference in favor of the PL and LPPL models against the benchmark model,
which supports the claim that the super-exponential property of the price
constitutes an important characteristics of financial bubbles.

\section{Out-of-sample tests of the volatility-confined LPPL model to diagnose other bubbles}

We now apply the above described procedure and tests of significance for the LPPL property to different price time series that contain  other historical speculative bubbles. Our goal is to test for the validity and universality of the volatility-confined LPPL model.

We proceed in two steps. For each time series to be analyzed, we first calibrate the nonlinear model 
(\ref{E: r_JLS}) with (\ref{E: landau 1}) to the logarithm of the price. If the 
LPPL parameters determined from the fit for a  certain period meet the LPPL conditions (\ref{E: LPPL condition}), a  speculative bubble is then diagnosed within this period. The  volatility-confined LPPL model is then supposed to be applicable. Second, we  test the O-U property as well as the order of autoregression of the residuals obtained 
from the previous calibration step  in the same time interval. This is a test of the stationarity of the 
residual time series.

We consider some of the most important speculative bubbles that have occurred in the World in the last decades.
Specifically, we study
\begin{itemize}
\item  the bubble  in the USA as well as in other European markets that led to a crash at the end of the summer of 1998 (the so-called Russian crisis), 
\item the booming market in Hong Kong in the mid-1990s ending with a crash of October 1997,
\item the ITC bubble reflecting over-optimistic expectation of a new economy ending in the spring of 2000 with a big crash of the NASDAQ index, 
\item the so-called oil bubble which started arguably around mid-2003 and ended in July 2008 \citep{Didier2009},
through it marks on the S\&P500 index and,
\item the recent Chinese bubble, characterized by crazy ups and downs and a sixfold increase 
of the Chinese indices in just two years, followed by a dramatic drop in a mere half year to one-third of its peak value attained in October 2007. We use the Shanghai Stock Exchange Composite index (SSEC) and Shenzhen Stock Component index (SZSC), which are two of the major stock index in Chinese market.
\end{itemize}

Table \ref{T:6} displays the parameters obtained from the calibration of the LPPL model to these bubbles.
One can verify that the LPPL conditions $B>0$, $0.1\le\beta \le 0.9$, $6\le\omega\le13$, and $|C|<1$ are met
for these bubbles\footnote{For the SSEC index, the estimated $\beta$ is found equal to $0.905$, which
is barely outside the chosen qualifying interval $[0.1,0.9]$. Changing slightly by a few days the time
window in which the fit is performed puts back the exponent $\beta$ within the qualifying interval.}.

Table \ref{T:7} gives the results of the O-U test for the residuals obtained from
calibrating the nonlinear model 
(\ref{E: r_JLS}) with (\ref{E: landau 1}) to the logarithm of each time series.
Combining the results of the different unit-root tests, we conclude that all indices except one
have their residuals qualifying as generating by a stationary process at the 
99.9\% confidence level. The exception is the Shenzhen stock component index for which
the confidence level to reject the null of non-stationarity is 99\%.
The estimated coefficient $\alpha$ of auto-regression associated
with the O-U process is between $0.02$ and $0.06$. This range of values corresponds 
approximately to our choice for the prior distribution of the coefficient $\alpha$ in the Bayesian analysis 
reported in the previous section. The last columns of Table\ref{T:7} list the order of the AR model 
obtained for the residuals. Two criteria of order selection are tested for robustness.
In almost all cases, the two different criteria give the same order equal to $1$ for the AR model, with only one exception being the Hang Seng Index for which the HQ criterion suggests an AR(3).

The above tests performed on these seven bubbles presented in Tables \ref{T:6}  and \ref{T:7} suggest that our
proposed volatility-confined LPPL model, first tested for the
bubble and crash of October 1987, is not just fitting a single ``story'' but 
provide a consistent universal description of financial bubbles, namely
a super-exponential acceleration of price decorated with log-periodic oscillations
with mean-reverting residuals.

\section{Concluding remarks}

We have presented a model of bubbles, termed the volatility-confined LPPL model,
to describe and diagnose
situations when excessive public expectations of future price increases 
cause prices to be temporarily elevated.

To break the stalemate in the literature concerning the detection of bubbles,
we have proposed to focus on three characteristics: (i) the faster-than-exponential
growth of the price of the asset under consideration represented
by a  singular power law behavior, (ii) an accelerated succession
of transient increases followed by corrections captured by a so-called
log-periodic component and (iii) a mean-reversing behavior of the residuals
developing around the two first components, which by themselves
form the log-periodic power law (LPPL) model. 

These three properties have been nicely tied together via a rational-expectation (RE) model
of bubbles with combined Wiener and Ornstein-Uhlenbeck innovations describing
the dynamics of rational traders coexisting with noise traders driving the crash hazard rate.
An alternative model has been proposed in terms of 
a behavioral specification of the dynamics of the 
stochastic discount factor describing the overall combined decisions of both rational
and noise traders. 

The test of the volatility-confined LPPL model has proceeded in two steps.
First, we calibrated the nonlinear model 
(\ref{E: r_JLS}) with (\ref{E: landau 1}) to the logarithm of the price time series
under study and diagnosed a bubble when the 
LPPL parameters determined from the fit for a  certain period meet the LPPL conditions (\ref{E: LPPL condition}).
Second, we tested for the stationarity of the residual time series.
Applied extensively to GARCH benchmarks and to eight historical well-known bubbles,
we found overall that these bubbles obey the conditions for the volatility-confined LPPL model
at a very high confidence level (99.9\%) and that the rate of false positives is very low, at
about $0.2\%$.  These results suggest that 
we have identified a consistent universal description of financial bubbles, namely
a super-exponential acceleration of price decorated with log-periodic oscillations
with mean-reverting residuals.

Further validation will come by testing further on other known bubble cases and in real time.
These studies are currently underway and will be reported elsewhere.

\bigskip
\noindent{\bf Acknowledgments}
\noindent 
The authors would like to thank Anders Johansen and Vladilen Pisarenko for useful discussions and
Ryan Woodard and Zhi-Qiang Jiang for help in the implementation of the tests. 
We acknowledge financial support  from the ETH Competence Center ``Coping with Crises in Complex 
Socio-Economic Systems'' (CCSS) through ETH Research  Grant CH1-01-08-2.
 This work was partly supported by the National Natural Science Foundation of China for Creative Research Group: {\em Modeling and Management for Several Complex Economic System Based on Behavior } (Grant No.\,70521001). Lin Li also appreciates the China Scholarship Council (CSC) for supporting his studies at ETH Zurich (No.\, 2008602049). 

\clearpage
\bibliographystyle{elsarticle-harv}
\bibliography{Critical_Bubbles_MeanReversing}

\pagebreak

\begin{table}[h]
\caption{Test of the LPPL specifications (\ref{E: landau 1})
to synthetic time series generated with the GARCH model (\ref{gthogrbrgbwq}), with the 
LPPL conditions (\ref{E: LPPL condition}), and the unit-root tests on the residuals. For each type
of samples, 1'000 time series have been generated.} \label{T: 1}
\begin{center}
\begin{tabular}{c  c   c c c  c }\hline\hline
type of &percentage of &signif. &\multicolumn{2}{c}{percentage of not rejecting $H_0$}&false positive \\
\cline{4-5}
samples & LPPL condition satisfied & level &Phillips-Perron&Dickey-Fuller&rate
\\\hline\hline
random & \multirow{3}*[2mm]{0.2\%}&$\alpha=0.01$&94.1\%&94.1\% &0.2\%\\
\cline{3-6}
length&&$\,\,\,\alpha=0.001$&72.8\%&72.8\%&0.2\%\\\hline
fixed & \multirow{3}*[2mm]{0.1\%}&$\alpha=0.01$&93.8\%&93.8\% &0.1\%\\
\cline{3-6}
length&&$\,\,\,\alpha=0.001$&72.7\%&72.7\%&0.0\%\\\hline\hline
\end{tabular}
\end{center}
\end{table}

\clearpage
\begin{table}[h]
\caption{Test of the LPPL specifications (\ref{E: landau 1})  and the unit-root tests on the residuals,
for time series of $750$ consecutive
trading days of the S\&P500 US index in the interval from Jan.~3, 1950 to Nov.~21, 2008. 
The first (respectively second) set of windows is obtained by sliding windows of $750$ days
over the whole duration of our data set with time increments of 
25 days (respectively 50 days). $P_{\mathrm{LPPL}}$ denotes the fraction of windows that satisfy the LPPL condition. $P_{\mathrm{Stationary Resi.}\mid \mathrm{LPPL}}$ is the conditional probability that, out of the fraction $P_{\mathrm{LPPL}}$ of windows that satisfy the LPPL condition, the null unit-root test for non-stationarity is rejected for the residuals.
}\label{T: 2}
\begin{center}
\begin{tabular}{c  cc   c c c  c }\hline\hline
days of & number of &\multirow{3}*[2mm]{$P_{\mathrm{LPPL}}$}&signif. &\multicolumn{2}{c}{percentage of not rejecting $H_0$}&\multirow{3}*[2mm]{$P_{\mathrm{Stationary Resi.}\mid \mathrm{LPPL}}$}\\
\cline{5-6}
one step &windows &  & level &Phillips-Perron&Dickey-Fuller&
\\\hline\hline
\multirow{3}*[2mm]{25} & \multirow{3}*[2mm]{563}&\multirow{3}*[2mm]{2.49\%}&$\alpha=0.01$&96.45\%&96.45\% &100\%\\
\cline{4-7}
& &&$\,\,\,\alpha=0.001$&69.27\%&69.27\%&100\%\\\hline
\multirow{3}*[2mm]{50}  &\multirow{3}*[2mm]{282} &\multirow{3}*[2mm]{2.84\%}&$\alpha=0.01$&96.81\%&96.81\% &100\%\\
\cline{4-7}
& &&$\,\,\,\alpha=0.001$&70.92\%&70.92\%&100\%\\\hline\hline
\end{tabular}
\end{center}
\end{table}

\clearpage
\begin{table}[h]
\caption{Windows of the S\&P500 US index in the interval from Jan.~3, 1950 to Nov.~21, 2008
that obey the LPPL conditions. Windows of type I (respectively type II)
are obtained by sliding a time interval of $750$ days
over the whole duration of our data sets with time increments of 
25 days (respectively 50 days).}\label{T: 3}
\begin{center}
\begin{tabular}{cccc}\hline\hline
start of window&end of window&reject $H_0$ for residuals& type of sliding step\\\hline\hline 
May.\,\,\,7, 1984&Apr. 24, 1987 & Yes &I\\
Jun. 12, 1984&Jun. \,\,\,1, 1987&Yes & I \& II\\
Jun. 18, 1984&Jul. \,\,\,\,7, 1987&Yes&I\\
Mar. 15, 1991&Feb. 16, 1994&Yes&I \& II\\
Mar. 25, 1994&Mar. 13, 1997&Yes&I\\
May.\,\,\,3, 1994&Apr. 18, 1997&Yes&I \& II\\
Jun. \,\,\,8, 1994&May. 23, 1997&Yes&I\\
Jul. 14, 1994 &Jun. 30, 1997&Yes&I \& II\\
Sep. 23, 1994&Sep. 10, 1997&Yes&I \& II\\
Oct. 28, 1994&Oct. 15, 1997&Yes&I\\
Apr. 28, 1995&Apr. 11, 1998&Yes&I \& II\\
Jun. \,\,\,5, 1995&May. 15, 1998&Yes&I \\
Jun. 11, 1995&Jun. 21, 1998&Yes&I \& II\\
Sep. 16, 1996&Sep. 30, 1999&Yes&I \& II\\\hline\hline
\end{tabular}
\end{center}
\label{default}
\end{table}%

\clearpage
\begin{table}[h]
\caption{Test for the validity of the LPPL conditions and unit-root tests on residuals in windows 
all ending on Sep.~30, 1987 with different starting dates for the S\&P500 US index. The smallest window size is 
750 days. $P_{\mathrm{LPPL}}$ is the percentage of windows that obey the LPPL conditions in all the test windows.  $P_{\mathrm{Stationary Resi.}\mid \mathrm{LPPL}}$ is the probability that
the null unit-root tests for non-stationarity are rejected for the residuals, conditional on the fact that the LPPL conditions are met. The unit-root tests are also the Phillips-Perron and Dickey-Fuller tests (both produce the same results) with significance level of $0.001$.}
\begin{center}
\begin{tabular}{ccccc}\hline\hline
start of    & number of & number of series          & \multirow{3}*[2mm]{$P_{\mathrm{LPPL}}$} & \multirow{3}*[2mm]{$P_{\mathrm{Stationary Resi.}\mid \mathrm{LPPL}}$}  \\
window & samples       & satisfy LPPL condition&&                                                                                        \\\hline\hline
Jan. 2, 1980 & 242 & 43 & 17.78\% & 100\%**\\
Jan. 3, 1983 & 90 & 43 & 47.48\% & 100\%**\\
Sep. 1, 1983 & 57 & 42 & 73.68\% & 100\%**\\
Dec. 1, 1983 & 44 & 43 & 97.73\% & 100\%**\\
Mar. 1, 1984 & 32 & 32 & \;\;100\% & 100\%**\\
\hline\hline
\end{tabular}
\end{center}
\label{T: 4}
\end{table}%

\clearpage
\begin{table}[h]
\caption{Phillips-Perron unit root test on residuals of the calibration of the S\&P 500 index by the LPPL model (\ref{E: r_JLS}) with (\ref{E: landau 1}) over the interval  from Jan.~3  1984 to Sep.~30 1987.}
\begin{center}
\begin{tabular}{ccccccccc}\hline\hline
&&&&&&Adj.~t-Stat&&Prob.*\\
\multicolumn{6}{l}{Phillips-Perron test statistic}&-4.008&&0.0001\\\hline
Test critical values&\multicolumn{5}{c}{0.1\%}&-3.588&&\\
  &\multicolumn{5}{c}{\,\,\,\,1\%}&-2.567&&\\
                                     &\multicolumn{5}{c}{\,\,\,\,5\%}&-1.941&&\\\hline\hline
     Model&&Coefficient $\alpha$&&Std.Error&&s-Statistic&&Prob.\\
     $\nu_{t+1}=-\alpha \nu_t+u_t$&&0.029&&0.0077&&-3.789&&0.0002\\\hline\hline
     R-squared&&0.015&&\multicolumn{3}{c}{Mean dependent var}&&-9.95E-05\\
     Adjusted R-sqaured&&0.015&&\multicolumn{3}{c}{S.D. dependent var}&&0.0084\\
     S.E. of regression&&0.0084&&\multicolumn{3}{c}{AIC}&&-6.7286\\
     Sum squared resid&&0.0662&&\multicolumn{3}{c}{SC}&&-6.7234\\
     Log likelihood&&3186.97&&\multicolumn{3}{c}{Durbin-Watson stat}&&1.7928\\\hline\hline
\end{tabular}
\end{center}
\label{T: 5}
\end{table}%

\clearpage
\begin{table}[h]
\caption{Parameters obtained from the calibration of the nonlinear model 
(\ref{E: r_JLS}) with (\ref{E: landau 1}) to the logarithm of the different price indices
named in the first column.} \label{T:6}
\begin{center}
\begin{tabular}{ccccccccc}\hline\hline
index&$t_{\mathrm{start}}$&$t_{end}$&$t_c$&$\beta$&$\omega$&$\phi$&B&C\\\hline\hline
S\&P500&Jan-03-91&Apr-30-98&Jul-11-98&0.3795&6.3787&4.3364&0.0833&0.7820\\
FTSE100&Jun-01-94&May-30-98&Aug-26-98&0.4022&12.1644&0.9409&0.0571&0.8076\\
HangSeng&Jan-03-95&Jul\,-31-97&Oct-28-97&0.7443&7.4117&4.9672&0.0042&0.7955\\
NASDAQ&Apr-01-97&Feb-28-00&May-27-00&0.1724&7.3788&3.2314&1.0134&0.9745\\
S\&P500&Dec-01-04&Jul-15-07&Oct-26-07&0.1811&12.9712&1.5361&0.2419&-0.8884\\
SSEC&Feb-01-06&Oct-31-07&Jan-23-08&0.9050&7.3538&2.3614&0.0054&-0.6277\\
SZSC&Feb-01-06&Oct-31-07&Dec-14-07&0.8259&6.3039&6.2832&0.0111&0.7344\\
\hline\hline
\end{tabular}
\end{center}
\end{table}%

\clearpage
\begin{table}[h]
\caption{Stationarity tests on the residuals of the financial indices obtained from the 
the calibration of the nonlinear model 
(\ref{E: r_JLS}) with (\ref{E: landau 1}) to the logarithm of the different price indices
named in the first column.  Tripled stars(***) and double stars(**) respectively denote 0.1\% and 1\% significance levels to reject the null $H_{0}$ that the residual process has a unit root. $\alpha$ is the mean-reverting parameter of the Ornstein-Uhnlenbeck generating process of the residuals. The orders of the AR model for the residuals selected using the Schwarz information Criterion (SIC) and the Hannan-Quinn Criterion are listed in the last two columns.}\label{T:7}
\begin{center}
\begin{tabular}{cccccccc}\hline\hline
\multirow{3}*[2mm]{index}&\multirow{3}*[2mm]{$t_{\mathrm{start}}$}&\multirow{3}*[2mm]{$t_{\mathrm{end}}$}&\multicolumn{2}{c}{unit-root test}&\multirow{3}*[2mm]{Coefficient $\alpha$}&\multicolumn{2}{c}{AR order}\\
\cline{4-5}\cline{7-8}
&&&Phillips-Perron&Dickey-Fuller&&SIC&HQ\\\hline\hline
S\&P500&Jan-03-91&Apr-30-98&-4.454$^{***}$&-4.594$^{***}$&0.022&1&1\\
FTSE100&Jun-01-94&May-30-98&-4.731$^{***}$&-4.893$^{***}$&0.045&1&1\\
HangSeng&Jan-03-95&Jul-31-97&-3.756$^{***}$&-3.482$^{***}$&0.041&1&3\\
NASDAQ&Apr-01-97&Feb-28-00&-3.849$^{***}$&-3.759$^{***}$&0.037&1&1\\
S\&P500&Dec-01-04&Jul-15-07&-4.000$^{***}$&-4.229$^{***}$&0.053&1&1\\
SSEC&Feb-01-06&Oct-31-07&-3.932$^{***}$&-3.808$^{***}$&0.064&1&1\\
SZSC&Feb-01-06&Oct-31-07&\!\!-3.111$^{**}$&\!\!-2.960$^{**}$&0.041&1&1\\
\hline\hline
\end{tabular}
\end{center}
\end{table}%

\clearpage 

\clearpage 

\end{document}